\documentclass[conference, 9pt]{IEEEtran}
\IEEEoverridecommandlockouts
\usepackage{cite}
\usepackage{amsmath,amssymb,amsfonts}
\usepackage{graphicx}
\usepackage{textcomp}
\usepackage[table,xcdraw]{xcolor}
\def\BibTeX{{\rm B\kern-.05em{\sc i\kern-.025em b}\kern-.08em
    T\kern-.1667em\lower.7ex\hbox{E}\kern-.125emX}}

\author{
Hanqing Zhu\textsuperscript{$\star$, 1},
Zhican Zhou\textsuperscript{2} ,
Shupeng Ning\textsuperscript{1} , 
Xuhao Wu\textsuperscript{2} , 
Ray T. Chen\textsuperscript{1} , 
Yating Wan\textsuperscript{2} , 
David Z. Pan\textsuperscript{$\dagger$,1}, 
\\
\textsuperscript{1} The University of Texas at Austin, \\
\textsuperscript{2} King Abdullah University of Science \& Technology \\
\textsuperscript{$\star$}hqzhu@utexas.edu; \textsuperscript{$\dagger$}dpan@ece.utexas.edu
}

\usepackage[utf8]{inputenc} %
\usepackage[T1]{fontenc}    %
\usepackage{pifont}

\usepackage{algorithm}
\usepackage{graphicx}
\usepackage{threeparttable}
\usepackage{multirow}
\usepackage{amsmath}
\usepackage{bm}
\usepackage{bbm}
\usepackage{amsthm}
\usepackage{soul}
\usepackage[noend]{algpseudocode}
\usepackage{enumitem} %
\usepackage[subrefformat=parens,labelformat=parens]{subfig}

\usepackage{wrapfig}
\usepackage{xspace}

\definecolor{citecolor}{RGB}{34,139,34}
\definecolor{mydarkblue}{rgb}{0,0.08,1}
\definecolor{mydarkgreen}{rgb}{0.02,0.6,0.02}
\definecolor{mydarkred}{rgb}{0.8,0.02,0.02}
\definecolor{mydarkorange}{rgb}{0.40,0.2,0.02}
\definecolor{mypurple}{RGB}{111,0,255}
\definecolor{myred}{rgb}{1.0,0.0,0.0}
\definecolor{mygold}{rgb}{0.75,0.6,0.12}
\definecolor{myblue}{rgb}{0,0.2,0.8}
\definecolor{mydarkgray}{rgb}{0.,0.2,0.2}

\definecolor{lightred}{RGB}{255,235,235}
\definecolor{lightgreen}{RGB}{235,255,235}
\definecolor{lightblue}{RGB}{235,235,255}
\definecolor{lightcyan}{RGB}{235,255,255}
\definecolor{lightmagenta}{RGB}{255,235,255}
\definecolor{lightyellow}{RGB}{255,255,235}

\definecolor{qxkcolor}{RGB}{215,235,255}
\definecolor{softmaxcolor}{RGB}{230,235,255}
\definecolor{probxvcolor}{RGB}{255,255,235}

\definecolor{topkcolor}{RGB}{255,235,235}
\definecolor{zecolor}{RGB}{255,255,235}
\definecolor{dynacolor}{RGB}{235,255,255}

\definecolor{reviewcolor}{RGB}{0,0,200}

\newcommand{\calO}{\mathcal{O}}

\newcommand{\calL}{\mathcal{L}}

\theoremstyle{plain}

\theoremstyle{definition}

\algdef{SE}[DOWHILE]{Do}{doWhile}{\algorithmicdo}[1]{\algorithmicwhile\ #1}%

\newcommand{\squishlist}{
 \begin{list}{$\bullet$}
  { \setlength{\itemsep}{0pt}
     \setlength{\parsep}{3pt}
     \setlength{\topsep}{3pt}
     \setlength{\partopsep}{0pt}
     \setlength{\leftmargin}{1.5em}
     \setlength{\labelwidth}{1em}
     \setlength{\labelsep}{0.5em} } }
     
\newcommand{\squishend}{
  \end{list}  }

\usepackage{amsmath,amsfonts,bm}

\def\eqref#1{equation~\ref{#1}}

\def\1{\bm{1}}

\def\vr{{\bm{r}}}
\def\vs{{\bm{s}}}

\def\mA{{\bm{A}}}
\def\mB{{\bm{B}}}

\def\mD{{\bm{D}}}

\def\mL{{\bm{L}}}

\def\mS{{\bm{S}}}

\def\mU{{\bm{U}}}
\def\mV{{\bm{V}}}
\def\mW{{\bm{W}}}
\def\mX{{\bm{X}}}

\DeclareMathAlphabet{\mathsfit}{\encodingdefault}{\sfdefault}{m}{sl}
\SetMathAlphabet{\mathsfit}{bold}{\encodingdefault}{\sfdefault}{bx}{n}

\newcommand{\R}{\mathbb{R}}

\DeclareMathOperator*{\argmin}{arg\,min}

\usepackage[colorlinks,linkcolor=red,citecolor=citecolor,urlcolor=blue]{hyperref}       % hyperlinks
\usepackage{graphicx}% Include figure files
\usepackage{dcolumn}% Align table columns on decimal point
\usepackage{bm}% bold math
\usepackage[utf8]{inputenc}
\usepackage[T1]{fontenc}
\usepackage{mathptmx}
\usepackage{etoolbox}

\usepackage{booktabs}
\usepackage{tabularray}
\usepackage{multirow}

\usepackage{amsmath}
\usepackage{amssymb}
\usepackage{amsthm}
\usepackage{fixmath}
\usepackage{amsmath}
\usepackage{amsfonts}
\usepackage{amssymb}
\definecolor{Highlight}{rgb}{0.94,1,0.94}
\definecolor{Highlight2}{rgb}{0.92,0.94,0.85}
\definecolor{Highlight3}{rgb}{0.82, 0.84, 0.75}
\usepackage{fontawesome}

\definecolor{gray1}{rgb}{0.6, 0.6, 0.6}
\definecolor{gray2}{rgb}{0.75, 0.8, 0.99}
\definecolor{gray3}{rgb}{0.84,0.89,0.99}
\definecolor{gray4}{rgb}{0.93,0.95,0.99}

\usepackage[breakable]{tcolorbox}
\DeclareRobustCommand{\questionbox}[2][pink!20]{%
\begin{tcolorbox}[   %% Adjust the following parameters at will.
        breakable,
        left=0pt,
        right=0pt,
        top=0pt,
        bottom=0pt,
        colback=#1,
        colframe=#1,
        width=\columnwidth,
        arc=0pt,outer arc=0pt,
        ]
        #2
\end{tcolorbox}
}

\newcommand{\name}{\texttt{ENLighten}\xspace}
\newcommand{\algname}{\texttt{Lighten}\xspace}
\newcommand{\algnamea}{\texttt{Lighten-I}\xspace}
\newcommand{\algnameb}{\texttt{Lighten-II}\xspace}

\newcommand{\baselinea}{\texttt{OATS}\xspace}

\makeatletter
\def\@email#1#2{%
 \endgroup
 \patchcmd{\titleblock@produce}
  {\frontmatter@RRAPformat}
  {\frontmatter@RRAPformat{\produce@RRAP{*#1\href{mailto:#2}{#2}}}\frontmatter@RRAPformat}
  {}{}
}%
\makeatother

\begin{document}

\title{
ENLighten: \underline{Lighten} the Transformer, \underline{En}able Efficient Optical Acceleration
}

\maketitle

\begin{abstract}
\label{abstract}
Photonic computing has emerged as a promising substrate for accelerating the dense linear‑algebra operations at the heart of AI, but its adoption for large Transformer models remains in its infancy.
In supporting these massive models, we identify two key bottlenecks:
(1) costly electro‑optic conversions and data‑movement overheads that erode energy efficiency as model sizes scale;
(2) a mismatch between limited on‑chip photonic resources and the scale of Transformer workloads, which forces frequent reuse of photonic tensor cores and dilutes throughput gains.
To address these challenges, we introduce a hardware–software co‑design framework. First, we propose \algname, a \textit{PTC‑aware compression flow} that post‑hoc decomposes each Transformer weight matrix into a low‑rank component plus a structured sparse component aligned to photonic tensor‑core granularity, all without lengthy retraining. Second, we present \name, a reconfigurable photonic accelerator architecture featuring dynamically adaptive tensor cores, driven by a broadband light redistribution, for fine‑grained sparsity support and full power gating of inactive parts. 
On ImageNet, \algname prunes Base-scale vision transformer by 50\% with a $\sim$1\% accuracy drop after only 3 epochs of fine-tuning within 1 hour,  and when deployed on \name it achieves a 2.5× improvement in energy–delay product over the state‑of‑the‑art photonic Transformer accelerator.

\end{abstract}
\section{Introduction}
\label{sec:Introduction}

As Moore’s Law winds down, optical computing has emerged as a promising new substrate for AI acceleration. 
By harnessing the inherent massive parallelism, low latency, and ultra-fast propagation of light, photonic tensor cores (PTCs) perform matrix–vector and matrix–matrix multiplications, via coherent interference~\cite{NP_NATURE2017_Shen, NP_Science2024_Xu, NP_HPCA2024_Zhu, NP_ACS2022_Feng, NP_NatureComm2022_Zhu} or incoherent multi‑wavelength intensity modulation and accumulation~\cite{NP_SciRep2017_Tait, NP_Nature2021_Xu, NP_Nature2021_Feldmann}.
These capabilities translate into orders‑of‑magnitude gains in throughput and energy efficiency over electronic accelerators for linear‑algebra workloads, making PTCs a natural match for the \textit{compute‑bound operations} that dominate many AI inference pipelines~\cite{touvron2022deit, ning2024photonic}.

Transformers have powered dramatic advances across both vision and language~\cite{hurst2024gpt,comanici2025gemini,agarwal2025cosmos,cong2025e3d,cong2025can}, yet their explosive growth, from hundreds of millions to billions of parameters, creates severe deployment burdens. In particular, non‑autoregressive architectures like Vision Transformers (ViTs)~\cite{touvron2022deit} rely heavily on large, batched matrix multiplications, a workload that is inherently compute‑bound and thus a perfect candidate for photonic acceleration. This raises a central question: \emph{can we leverage photonic tensor cores to unlock new levels of speed and efficiency for large‑scale Transformer inference?}

\begin{figure}
    \centering
    \includegraphics[width=\columnwidth]{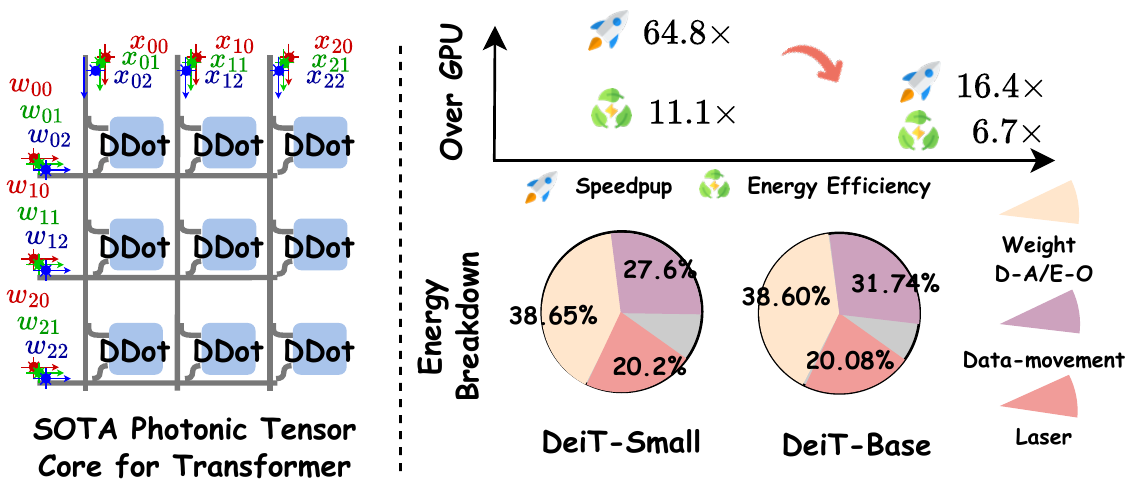}
    \vspace{-10pt}
    \caption{(left) State-of-the-art PTC design in photonic Transformer accelerator; (right) their diminishing benefits on speedup and energy efficiency moving from DeiT-Small to DeiT-Base; their energy cost is bottlenecked by data movements and weight digital-to-analog and modulation cost.}
    \vspace{-10pt}
    \label{fig:Teaser}
    \vspace{-5pt}
\end{figure}

Despite widespread demonstrations of PTCs for convolutional networks, applying PTCs to medium‑ and large‑scale Transformers remains in its infancy.
The recent photonic Transformer accelerators~\cite{NP_HPCA2024_Zhu} fill this gap with impressive speedups and energy efficiency over digital counterparts, shown in Fig~\ref{fig:Teaser}.
However, these benefits markedly diminish as model size grows: throughput improvements drop from hundreds of $\times$ to only tens of $\times$, and energy‑efficiency gains fall from $\sim$30× to under 10× when scaling from ViT‑Tiny to ViT‑Base on their 8-bit version. We trace this erosion to two fundamental bottlenecks:
(1) \textbf{Electro‑optic conversion and data‑movement overhead.} Converting between electrical and optical domains and shuttling large weight tensors incur substantial energy and latency costs, often eroding more than half of the energy cost.
(2) \textbf{Photonic chip scale vs.\ model scale.} Bulky photonic components and mature fabrication nodes limit on‑chip PTC density, forcing time‑multiplexed reuse of a small pool of cores and diluting end‑to‑end throughput as models grow.

These observations motivate a hardware–software co‑design approach: \emph{compress} large Transformer models so that the reduced scale of the model better matches the strengths and constraints of photonic accelerators. While digital model‑compression techniques have achieved impressive parameter reductions in electronic systems~\cite{zhang2025oats, sun2024a}, naively porting them to optics fails to translate into real energy or latency gains due to the need for PTC‑friendly, structured sparsity patterns and granularity.
Prior optical pruning efforts have explored phase‑shifter removal in MZI arrays or PTC‑aware row/column sparsity (e.g., SCATTER~\cite{yin2024scatter}), yet they either incur no throughput speedup or suffer large accuracy drops (e.g., > 6\% on CIFAR‑100) even after extensive sparse training. Such approaches are impractical for Transformers, since retraining them demands prohibitive computational resources.

Given this context, we pose the central question of this paper:
\questionbox{
\textit{\textbf{Question}:}~\textit{Can we \underline{compress large Transformer models} to improve \underline{hardware efficiency} on photonic AI engines, while incurring \underline{minimal accuracy loss} under \underline{limited retraining}?}
}

\medskip\noindent Our answer is two‑fold: \ding{202} We present \algname, a \emph{PTC‑aware compression flow} that decomposes each weight matrix into a dense low‑rank term and a PTC‑aligned structured sparse term. Through activation‑aware decomposition, fast rank allocation, and a two‑stage distillation, we preserve fidelity with only a handful of fine‑tuning epochs (e.g., we set the fine-tuning time budget to 1 hour, which is 3 epochs for ViT-Base).
\ding{203} We co‑design \name, a photonic accelerator that augments a standard dense optical engine with a reconfigurable sparse engine capable of exploiting the column‑wise sparsity produced by \algname, delivering genuine energy savings and throughput speedup.

The major contributions of this paper are as follows:

\squishlist \item \textbf{\algname:} the first post‑hoc compression pipeline tailored to PTC granularity, combining activation‑aware low‑rank factorization with structured sparsity under minimal retraining.
\item \textbf{\name:} a reconfigurable photonic architecture featuring reconfigurable sparse engines with calibration‑free broadband switching with an adaptive operating dimension to execute the compressed models efficiently under different sparse granularities.
\item \textbf{End‑to‑end evaluation:} up to 50\% parameter reduction on base-scale Vision Transformers with $\sim$ 1\% accuracy loss, yielding up to 2.5× improvement in energy–delay product over the prior state‑of‑the‑art photonic Transformer accelerator. \squishend

\section{Background}
\label{sec:Background}
\subsection{Optical Computing Basics}
\label{sec:PhotonicsBasics}

Various photonic neural network designs encode inputs and weights to light magnitude/phase and circuit transmission, performing ultra-fast matrix multiplication based on either standard devices~\cite{NP_NATURE2017_Shen, zhu2022elight_aspdac, NP_PIEEE2020_Cheng, NP_NaturePhotonics2021_Shastri, NP_ACS2022_Feng, NP_NatureComm2022_Zhu,NP_Science2024_Xu, NP_HPCA2024_Zhu, zhu2022elight,NP_SciRep2017_Tait, NP_Nature2021_Xu, NP_Nature2021_Feldmann, zhu2022fuse} or customized devices~\cite{NP_NatureComm2022_Zhu, gu2022adept, gu2022neurolight, zhu2024pace, NP_TCAD2022_Gu, gu2024m3icro}.

Notably, recently, researchers have proposed an optical transformer accelerator, called \texttt{Lightning-Transformer}, short for \texttt{LT} in our paper.
A key enabling building block is the dynamically‑operated photonic tensor core (DPTC), which arranges many optical dot‑product (DDot) engines into a compact crossbar array to realize general matrix–matrix multiplication (GEMM).  Each DPTC has dimensions \(N_v\times N_h\), corresponding to \(N_v\) input waveguides in the “vertical” direction and \(N_h\) in the “horizontal” direction.  Within each waveguide, a WDM demultiplexer splits the incoming light into \(N_{\lambda}\) wavelength channels, each of which is modulated by a high‑speed Mach–Zehnder modulator (MZM) with the corresponding input or weight value.  These modulated signals are then recombined by a WDM multiplexer and fed into an array of directional couplers with appropriate phase shifts, causing the encoded amplitudes to interfere and produce photocurrents proportional to the inner products of input and weight vectors.

In one cycle, the DPTC thus implements an \([N_h \times N_{\lambda}]\times[N_{\lambda}\times N_v]\) matrix multiplication:  
\[
\mathbf{Y} = \mathbf{W}\,\mathbf{X},\quad
\mathbf{W}\in\R^{N_h\times N_{\lambda}},\;
\mathbf{X}\in\R^{N_{\lambda}\times N_v}.
\]
By sharing input and weight modulations across many DDot units within the core, DPTCs minimize the costly optical modulation overhead and fully exploit photonic parallelism, making them a powerful primitive for accelerating large Transformer workloads on optical hardware~\cite{NP_HPCA2024_Zhu}.

\subsection{Low-Rank Compression}
Modern neural networks, especially large Transformers, are massive, creating storage and compute bottlenecks. 
Alongside pruning~\cite{sun2024a,yin2024scatter} and quantization~\cite{liu2024spinquant,NP_DATE2020_Gu}, \emph{low-rank} compression exploits the empirical low-rank structure of weight matrices by retaining only the top-$k$ components. 
Low-rank parameterization has also proven effective for memory-efficient training~\cite{zhao2024galore,zhu2024apollo}. 
Beyond naive SVD truncation, \cite{yuan2023asvd} propose a data-aware factorization that directly minimizes activation error, improving fidelity at the same rank. 

Despite these advances, pure low-rank models often lag strong pruning/quantization baselines at comparable compression ratios. 
A promising remedy is to \emph{augment} low-rank factors with a small sparse residual~\cite{zhang2025oats}. However, most residuals are \emph{unstructured}, limiting realizable speedups on actual hardware.

\textbf{Our goal.}
We introduce, to our knowledge, the first compression toolkit that couples low-rank factors with \emph{hardware-aware} sparsity, i.e., structured patterns that map efficiently to photonic neuromorphic hardware, and integrates natively with its low-precision.
This low-rank\,+\,structured-sparsity\,+\,quantization trifecta is designed to deliver both accuracy retention and \emph{real} end-to-end acceleration.
\section{Proposed \name: Overview \& Problem}
\label{sec:Method}

In this paper, we investigate the following question:
\questionbox{
\textit{\textbf{Question}:}~\textit{Can we \underline{compress large Transformer models} to enhance its \underline{hardware efficiency} on photonic AI engines, while incurring \underline{minimal accuracy loss} under \underline{limited retraining}?}
}

This question is motivated by several key challenges:
\ding{202}\emph{Photonic chip scale vs. model scale:} 
Photonic chips remain limited in capacity due to their large footprint and mature fabrication processes. Even with light‑speed operation, they struggle to keep pace with the rapid growth of modern AI model sizes.
\ding{203} \emph{Energy inefficiency for large Transformers:}
Transformers incur significant data‑movement and analog$\leftrightarrow$digital conversion costs when mapped to photonic tensor cores, due to their high parameter counts, even with SOTA dedicated photonic accelerators~\cite{NP_HPCA2024_Zhu}.
\ding{204} \textit{High re-training cost}: 
Fully retraining large models is prohibitively expensive—for example, training ViT‑Base~\cite{touvron2022deit} on ImageNet for 300 epochs requires 8 GPUs for $\sim$ 2 days, making a compression approach with low retraining highly desirable.

\subsection{Low Rank plus Sparse Factorization}
We leverage the insight that a matrix \(\mathbf{W}\in\mathbb{R}^{m\times n}\) can be more faithfully approximated by the sum of a low‑rank component and a sparse component than by either alone, known as Robust PCA~\cite{candes2011robust}. This approach has proven effective for compressing both attention~\cite{chen2021scatterbrain} and linear layers~\cite{zhang2025oats}. Formally, we solve
\begin{equation}\label{eqn:rpca}
\small
    \begin{aligned}
        \min_{\mS,\, \mL} \quad & \| \mW - \mS - \mL \|_F^2 \\
        \text{s.t.} \quad & \text{Rank}(\mL) \leq r, \quad \|\mS\|_0 \leq s,
    \end{aligned}
\end{equation}
where \(\mL = \mA\mB, \mA \in \mathbb{R}^{m \times r}, \mB \in \mathbb{R}^{r \times n}\) represents the low-rank component and \(\mS\) denotes the sparse component with at most \(s\) nonzeros.

\definecolor{color1}{HTML}{FFCCCC}   % Light red
\definecolor{color2}{HTML}{CCCCFF}   % Light blue
\definecolor{color3}{HTML}{CCFFCC}   % Light green

\subsection{Our Problem Formulation and Solution}
We build upon the above compression idea, but extend it to model-level optimization, focusing beyond single matrix optimization.

\underline{\textbf{\emph{Problem Formulation:}}} 
Let \(\Theta\) be a Transformer with \(L\) layers and weight matrices \(\{\mW_l\}_{l=1}^L\). Given a global compression target \(\alpha\in(0,1)\), our goal is to compute per‑layer decompositions $
\mW_l \approx \mS_l + \mL_l$
where \(\mathrm{rank}(\mL_l)\le r_l\) and \(\|\mS_l\|_0\le s_l\), such that the overall parameter reduction
\(\psi(\{s_l,r_l\})\ge\alpha\).  
Limited fine‑tuning is conducted further.

Directly tackling this is difficult because several sub‑problems are tightly intertwined.
First, determining the per-layer budgets $(s_l, r_l)$ that satisfy the global compression target is an \emph{NP-hard combinatorial task}; these budgets, in turn, dictate each layer’s decomposition configuration.
Second, each layer must be faithfully decomposed into low‑rank and sparse terms; poor decompositions propagate error and make subsequent fine‑tuning far less effective.

\underline{\textbf{\emph{Our Solution:}}}
We decouple the joint objective into the three‑level hierarchy below, turning an intractable problem into three tractable sub‑tasks:
\begin{subequations}\label{eqn:problem.}
\small
    \begin{align}
        &\begin{aligned}
            &\colorbox{color1}{\texttt{Level 3:} \textit{Model Finetuning}} \\
            &\quad \Theta^*(\vs^*,\vr^*) = \argmin_{\Theta}\; \calL\Bigl(\Theta(\vs^*,\vr^*),\mathcal{D}_{task}\Bigr), \\
        \end{aligned} \label{eq:level3} \\
        &\begin{aligned}
            &\colorbox{color2}{\texttt{Level 2:} \textit{Model-wise Compression Setting Search}} \\
        &\quad (\vs^*,\vr^*) = \argmin_{\vs, \vr}\; \calL\Bigl(\Theta(\vs,\vr)\Bigr), \quad \\
        &\quad \text{s.t.} \quad \psi(\vs,\vr) \geq \alpha, \\[1mm]
        \end{aligned} \label{eq:level2} \\
        &\begin{aligned}
            &\colorbox{color3}{\texttt{Level 1:} \textit{Layer-wise Optimization}} \\
            &\quad \mS_{l}^{*}, \mL_{l}^{*} = \argmin_{\mW_l \in \Theta}\; \| \mW_l - \mS_l - \mL_l \|_{F}^2, \quad \forall l = 1, \ldots, L, \\
            &\quad \text{s.t.} \quad \text{Rank}(\mL_l) \leq \vr^*_l, \quad \|\mS_l\|_0 \leq \vs^*_l.
        \end{aligned} \label{eq:level1}
    \end{align}
\end{subequations}

\underline{Level 1 (Eq.~\ref{eq:level1}):}  
Given a layer‑specific budget $s_l, r_l$, we factorize each weight matrix while strictly adhering to the PTC‑aligned structured sparsity pattern (see Sec.\ref{subsec:sparse_pattern} for the sparsity discussion).
In Sec.~\ref{subsec:level1}, we introduce an activation‑aware decomposition with a local low‑rank adaptation to maximize layer‑wise fidelity. 

\underline{Level 2 (Eq.~\ref{eq:level2}):}  
We efficiently search for per‑layer compression settings \((\vs,\vr)\) with a fast, heuristic rank allocator in Sec.~\ref{subsec:level2}. 
Unlike uniform schemes (e.g., \baselinea), we adjust compression budget across layers by varying the ranks $r$ to exploit Transformers’ inherent diverse low‑rank weights~\cite{jaiswal2024galore}. \textit{The sparse ratio is fixed and small, since dense matrix from low-rank decomposition is more desired for optics.}
% In Sec.~\ref{subsec:level2}, we propose a fast, heuristic rank allocator that uses layer‑wise decomposition error as a guiding signal to redistribute the compression budget.

\underline{Level 3 (Eq.~\ref{eq:level3}):}  
We fine‑tune the compressed model on the target dataset for a handful of epochs (e.g., 3) to recover accuracy. Sec.~\ref{subsec:level3} presents a two‑stage distillation workflow, first aligning block‑wise outputs and then self‑teaching with the original model, to maximize performance under strict retraining constraints.

\begin{figure*}
    \centering
    \includegraphics[width=0.83\textwidth]{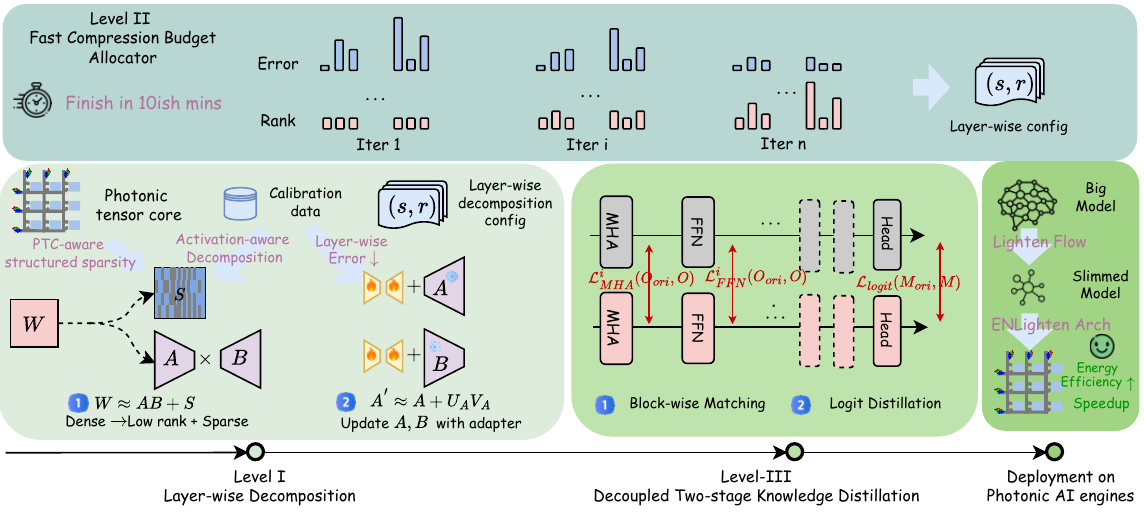}
    \caption{Overview of the \algname flow to trim a big model to a slimmed one for energy efficiency and speedup on Photonic AI engines.}
    \label{fig:sparseLT-alg}
\end{figure*}

\section{Algorithm: \algname, PTC-aware Compression Flow}
\label{sec:Method_alg}

To solve this three‑level problem, we propose \algname, a \textit{efficient} and \textit{high-fidelity} PTC‑aware compression pipeline that embeds hardware constraints into optimization, as illustrated in Fig.~\ref{fig:sparseLT-alg}.

\subsection{Level 1: Layer-wise Decomposition with Activation-awareness and Quick Low-Rank Adaptation}
\label{subsec:level1}
\underline{\textit{\textbf{Motivation:}}}
The standard layer-wise decomposition objective (Eq.~\ref{eq:level1}) implicitly assumes uniform input activations ($\mX_l = \mathbf{I}$), thereby neglecting the variations and prominent outliers often present in Transformer activation distributions~\cite{sun2024a}. Failing to account for these activation dynamics can lead to suboptimal compression.

\underline{\textit{\textbf{Our Solution:}}}
To address this, we formulate an \textit{activation-aware decomposition} objective that minimizes the reconstruction error with respect to the layer's input activations $\mX_l$:
\begin{equation}
\small
\begin{aligned}
\label{eqn:arpca}
  &\quad \mS_{l}^{*}, \mL_{l}^{*} = \argmin_{\mW_l \in \Theta}\; E_{\mX_l} \| \mW_l \mX_l - (\mS_l + \mL_l) \mX_l \|_{F}^2, \quad \forall l = 1, \ldots, L, \\
            &\quad \text{s.t.} \quad \text{Rank}(\mL_l) \leq \vr^*_l, \quad \|\mS_l\|_0 \leq \vs^*_l.
\end{aligned}
\end{equation}
% Here, the error minimization is performed with respect to the layer's input activations $\mX_l$.

Directly optimizing analytically Eq.~\ref{eqn:arpca} is challenging because the expectation operator $E_{\mX_l}$ requires integrating over all possible inputs.
We follow prior work~\cite{sun2024a, zhang2025oats} to capture the activation statistics with the second moment of inputs on a small calibration dataset $\mathcal{D}_{calib}$ (e.g., 2k samples). For each layer $l$, we compute a diagonal scaling matrix $\mD_l$ :
\begin{equation}
\small
\mD_l = \sqrt{\text{diag}(\mX_{calib, l}^{\top} \mX_{calib, l})},
\end{equation}
where $\mX_{calib, l}$ represents the input activations for layer $l$.

Using $\mD_l$, we approximate the activation-aware objective (Eq.~\ref{eqn:arpca}) with the following weighted reconstruction problem:
\begin{equation}
\small
\begin{aligned}
\label{eqn:arpca_v2}
  \mS_{l}^{*},\, \mL_{l}^{*} &= \argmin_{\mW_l \in \Theta,\; \mD_l}\; \| \big(\mW_l \mD_l - (\mS_l + \mL_l)\big)\mD_l^{-1} \|_{F}^2,\quad \forall l=1,\ldots,L,\\[1mm]
  &\quad \text{s.t. } \text{Rank}(\mL_l) \le \vr^*_l,\; \|\mS_l\|_0 \le \vs^*_l,
\end{aligned}
\end{equation}
which minimizes the reconstruction error of the scaled weight matrix $\mW_l \mD_l$. It can be solved by alternating optimization techniques in \baselinea~\cite{zhang2025oats}. Specifically, we iteratively (1) update the low-rank component $\mL_l$ using truncated SVD on the residual $(\mW_l\mD_l - \mS_l)$; 
(2) update the sparse component $\mS_l$ by applying hardware-aware structured sparsification to the residual $(\mW_l\mD_l - \mL_l)$.

\textbf{\textit{Key Design \ding{172}: Hardware-aware Structured Sparsification:}}
PTC-aware constraints have to be forced on the sparse pattern of the sparse term $\mS_l$ for future enhanced performance and speedup.
We discuss the PTC-friendly sparse pattern in Sec.~\ref{subsec:sparse_pattern}.
For a PTC in the size of $N_h, N_{\lambda}$, 
we first devide the weight block $\mW_l \in \mathbb{R}^{m\times n}$ into 3-d tensors $\mW_l \in \mathbb{R}^{\lceil \frac{m}{N_h}\rceil\times N_h \times n}$.
Then for each weight chunk $\mW_L[i, :, :]$, we prune column-wise vectors in the length of $N_h$ with a fixed sparse ratio $s$ (e.g., 10\%) based on their L1 norm~\cite{sun2024a}, resulting in $d=n\times s$ remaining nonzero entries.
Each weight chunk is pruned with the same sparsity ratio, therefore, can be further condensed to form a small sub-weight matrix, which can be mapped and accelerated on our dedicated sparse engine design discussed in Sec.~\ref{sec:Method_arch}.

\textit{\textbf{Key Design \ding{173}: Boost Fidelity via Local Low-Rank Adaptation:}}
While alternating optimization effectively minimizes the decomposition error, it would inevitably discard fine‑grained information. 
To address this, we introduce a fast \textit{local low‑rank adaptation} process using the same calibration dataset. After obtaining the initial factors \(\mA_l\) and \(\mB_l\), we initialize lightweight adapters \(\Delta\mA_l = \mU_{A_l}\mV_{A_l}\) and \(\Delta\mB_l = \mU_{B_l}\mV_{B_l}\), each of rank at most \(r_l/4\), and solve:
\begin{equation}
\small
\label{eq:local_regression}
\min_{X_i \in \mathcal{D}_{calib}} \; \Bigl\| \mW_l X_i - \Bigl[\bigl(\mA_l + \Delta \mA_l\bigr) \bigl(\mB_l + \Delta \mB_l\bigr) + \mS_l\Bigr] X_i \Bigr\|_F^2,
\end{equation}
The refined factors \(\Delta\mA_l\) and \(\Delta\mB_l\) can be merged back into the \(\mA_l\) and \(\mB_l\), incurring no additional parameters cost. This adaptation with small-rank adapter, requiring only a few gradient descent steps, significantly improves zero‑shot compression accuracy without overfitting on calibration dateset with the usage of small rank (See Tab.~\ref{tab:ModelAblation}).

\begin{algorithm}[tb]
    \caption{Fast Batch-wise Rank Allocator}
    \label{alg:rank_assignment}
    \small{
    \begin{algorithmic}[1]
        \Require Model $\mathcal{M}$, target compression ratio $\alpha$, initial rank list $\mathbf{r}^{init}$, fixed sparse ratios $\mathbf{s}$, $\tau$ step scheduling method, basis step sizes $b$, temperature $T$,
        \Ensure List of ranks $\mathbold{r}^{final}$
        \State \textcolor{gray}{Pre-decompose once with $(r_l,s_l)$ to get $\mA_l,\mB_l,\mS_l$ and error lists $\mathbf{e}$}
        \State Target parameters $p_{\mathrm{tar}} = \alpha \sum_{l} n_l \times d_l$
        \State Current parameters $p_{\mathrm{cur}} = \sum_{l}\big( r_l\,(n_l + d_l) + n_l\times d_l \times s_l \big)$
        \State Budget $B \gets p_{\mathrm{tar}} - p_{\mathrm{cur}}$
        % \For{each layer $l \in \{1,\ldots,L\}$}
        %     \State Obtain $\mA_l, \mB_l, \mS_l \leftarrow \mW_l$ \Comment{}
        %     \State Compute layer-wise error $e_l$ 
        % \EndFor
        \While{$B > 0$}
            % \State \textcolor{gray}{\# Step 1: Layer-wise Error Indicator.}
            \State \textcolor{gray}{\# Batch-wise Greedy Selection.}
            \State $e_l^{'} \gets e_l / \bigl(\sum_{j=1}^L e_l \bigr)\quad\forall l$ \Comment{\textcolor{gray}{normalize raw errors}}
            \State $\mathbf{P} \gets \mathrm{SoftMax}\!\bigl(\mathbf{e^{'}}/T\bigr)$
            % \State $\displaystyle Z \gets \sum_{j=1}^L \exp(e_j / T)$ \Comment{partition function}
            % \For{$l=1,\dots,L$}
            %     \State $P_l \gets \exp(e_l / T)\,/\,Z$     \Comment{soft‑max probability}
            % \EndFor
            \State Select minimal set $L_{\text{sel}}$ s.t. $\sum_{l\in L_{\text{sel}}}P(l) \ge p$
            \State \textcolor{gray}{\# Dynamic Step-Wise Refinement.}
            \State Re-normalize probabilities for $l\in L_{\text{sel}}$: $Q(l)$ with $\sum_{l\in L_{\text{sel}}} Q(l) = 1$
            \State $\Delta R = len(L_{\text{sel}}) \tau (B, p_cur, b)$
            % \State \textcolor{gray}{\# Step 4: Rank Update with Budget Check and Rejection $L_{\text{sel}}$.}
            \For{each $l\in L_{\text{sel}}$}
                \State $\Delta r_l \gets \Delta R \times Q_l$ \Comment{Assign increment rank based on $Q_l$}
                \State Update: $r_l \gets r_l + \Delta r_l$
                \State $\Delta p_l \gets \mathrm{ComputeParamIncrement}(l, \Delta r_l)$
                \If{\( \mathrm{LayerParams}(l, r_l, s_l)> \mathrm{OriLayerParams}(l) \)}
                    \State Set error: \( e_l \gets 0 \)
                    \State \textbf{continue} \Comment{Compression rejection for this layer}
                \EndIf
                \State Update error $e_l$ with new rank $r_l$
                \State $B \gets p_{\text{target}} - \Delta_l$
                \If{$B < 0$}
                    \State \textbf{break} \Comment{Terminate if target is exceeded}
                \EndIf
            \EndFor
        \EndWhile
        \State \Return $\mathbf{r}^{final} = \{r_l\}_{l=1}^L$
    \end{algorithmic}
    }
\end{algorithm}

\subsection{Level 2: Balancing Compression Error via Fast Batch-wise Rank Allocator}
\label{subsec:level2}

\underline{\textit{\textbf{Motivation:}}}  
Our closest baseline, OATS~\cite{zhang2025oats}, 
assigns the \emph{same} compression ratio to every layer, overlooking the fact that Transformer layers possess very different amounts of redundancy, which leads to highly uneven reconstruction error.
Searching for an per‑layer ranks is NP‑hard, since the search space grows as \(\calO\bigl(\prod_{l=1}^L d_l\bigr)\), where \(d_l\) is the maximum rank of layer \(l\).
Existing methods that tackle this search explicitly 
with the resource-intensive operations~\cite{yuan2023asvd, yu2023svd}, liking formatting as neural network search problem~\cite{yu2023svd}, nullifying the inference-time compression benefit in level I.
We therefore seek a lightweight, heuristic allocator that re‑distributes the rank budget non‑uniformly across layers.

\underline{\textit{\textbf{Our Solution:}}} 
We introduce a \textit{fast batch‑wise greedy rank allocator} (Alg.~\ref{alg:rank_assignment}) that, given a sparsity \(\vs\) and overall target \(\alpha\), produces per‑layer ranks \(\vr\) with minimal overhead (in 10ish minutes).
Sparsity itself is \emph{not} searched: dense low‑rank factors are far more photonic‑friendly, and Transformer layers are known to be intrinsically low‑rank \cite{jaiswal2024galore}.
The algorithm starts from a small uniform rank (e.g., 10\% of every layer’s maximum) and iteratively raises the ranks of the most error‑sensitive layers until the global budget is fully allocated.

\textbf{\textit{Key Design \ding{172}: Layer‑wise Error Indictor.}}
We score the compression quality of each layer with a normalized, activation‑aware reconstruction error,
\begin{equation}
\small
     e_l = \frac{\|\,[\mW_l\mD_l - (\mA_l\mB_l + \mS_l)]\\|_F}{\|\mW_l\mD_l\|_F},
\end{equation}
where \(\mD_l\) is the diagonal scaling matrix from Sec.~\ref{subsec:level1}.
Normalization by \(\|\mW_l\mD_l\|_F\) ensures comparability across layers. 
This choice has two advantages. (1) Since $\mD_l$ embeds activation statistics, the metric tracks true compression fidelity more reliably than computing MSE on extra calibration samples as we found empirically.
(2) It is cheap to use during the search: as ranks $r_l$ updates we need only reuse the stored full‑rank factors $\mA_l, \mB_l$, fetching the first $r$ columns/rows ($\mA_l[:,:r], \mB_l[:r,:]$); no additional SVD or data passes are required.

\textbf{\textit{Key Design \ding{173}: Batch‑wise Greedy Selection.}} 
To speed up the search, we operate on \emph{batches} of layers rather than updating one at a time.
At each iteration, we convert error list \(\{e_l\}\) into a soft‑max distribution and select a batch of layers in descending probability whose cumulative probability reaches a preset threshold (e.g., 50\%). Each selected layer’s rank is then increased by a step size \(\Delta r_l\).

% \hz{why we use basis rank, diferent layers may have differen params}

\textbf{\textit{Key Design \ding{174}: Dynamic Step‑wise Refinement.}}  
We tie the rank increment \(\Delta r\) to a “basis rank” \(b\), derived from the PTC core dimension \(P\) and the model’s hidden size (e.g., \(b=P=12\) for ViT‑Base, \(b=P/2=6\) for ViT-small). Rather than a fixed step, we use a linear decay schedule under a total params budget \(B\): 
\begin{equation}
\small
    \Delta r = 
\begin{cases}
2b, & \text{if remaining budget} \ge \tfrac{1}{2}B,\\
b,  & \text{if } \tfrac{1}{4}B \le \text{remaining budget} < \tfrac{1}{2}B,\\
\tfrac{b}{2}, & \text{otherwise.}
\end{cases}
\end{equation}
Instead of using the same $\Delta r$ for all selected layers, we compute the total rank budget for the batch $\Delta R \;=\; |L_{\text{sel}}| \times \Delta r$ and then redistribute it proportionally based on their probability.
Thus, layers with higher error probability receive a larger share of the
rank increment.
\textit{This coarse-to-fine manner leverages the singular-value spectrum: large early jumps eliminate the bulk of the error, while finer later updates capture diminishing-return components.}

\underline{\textit{\textbf{Efficiency:}}}  
The allocator is lightweight as (i) its coarse‑to‑fine schedule reduces the number of iterations, and (ii) all decomposition are performed on the \emph{already‑stored} full‑rank factors $\mA_{fr}, \mB_{fr}$.
No additional SVDs are required:  the new rank one is obtained by simple column/row slicing of $\mA_{fr}, \mB_{fr}$.
In practice, the entire rank search for a medium‑scale ViT‑Base completes in 10ish minutes on a single GPU, making it fast enough to \textit{run at deployment time.}

\subsection{Level 3: Boosting Accuracy via Decoupled Two-Stage Distillation}
\label{subsec:level3}

\underline{\textit{\textbf{Motivation:}}}  
Finally, we aim to recover accuracy via limited fine-tuning. 
We therefore require a compact training recipe that maximizes performance recovery within a very limited number of epochs.

\underline{\textit{\textbf{Our Solution:}}}  
We employ a decoupled two‑stage \emph{distillation}:

\textbf{\textit{Key Design \ding{172}: Block‑wise Error Alignment}}
This stage is motivated by the fact that Transformers are composed of \emph{blocks} (e.g., attention + MLP), and previous layer-wise decomposition ignores the connection.
Aware of this, the first stage focuses on aligning block outputs.
\begin{equation}
\small
    \mathcal{L}_{\text{block}} = \frac{1}{B}\sum_{b=1}^B \bigl\|F_b^S - F_b^T\bigr\|_2^2, \quad \text{block} \in [\text{Attn, MLP}]
\end{equation}
so that the student learns to mimic the teacher’s intermediate representations, reducing representational drift.

\textbf{\textit{Key Design \ding{173}: Logit‑Level Distillation}}  
In Stage 2, we apply logit distillation on the final outputs to recover task accuracy. Given student logits \(y^S\), teacher logits \(y^T\), and temperature \(\tau\), we optimize:
\begin{equation}
\small
\label{eq:LogitDistill}
\mathcal{L}_{\text{logit}}
= \tfrac{1}{2}\,\mathrm{KL}\bigl(y^S/\tau\,\|\,y^T/\tau\bigr)
+ \tfrac{1}{2}\,\mathrm{CE}\bigl(y^S,\,y_{\text{true}}\bigr),
\end{equation}
where \(\mathrm{KL}\) and \(\mathrm{CE}\) is the KL divergence and the cross‑entropy loss.

This two-stage process, first aligning block-level features, then final logits, maximizes accuracy recovery under a strict fine-tuning budget.

% \section{Proposed \name: Architecture }
\section{Hardware: \name, enable \algname on optics}
\label{sec:Method_arch}
In this section, we introduce \name, a reconfigurable photonic accelerator architecture that natively supports both dense matrix multiplications, on original weights or on the low‑rank factors produced by our \algname compression flow, as well as the structured sparse workloads it generates. 
To handle these sparse patterns efficiently, \name devises dynamically reconfigurable photonic tensor cores that can switch between multiple operating granulariy, enabling real energy saving by gating inactive parts and fine-grained processing where accuracy demands.

\subsection{PTC-aware Sparse Pattern}
\label{subsec:sparse_pattern}
We first review how weights are mapped onto the photonic engines~\cite{NP_HPCA2024_Zhu}.
Each PTC supports weight sub‑blocks of size \(N_v\times N_h\).
Given a linear layer with weight matrix \(\mW\in\R^{m\times n}\), we tile \(\mW\) into 
\begin{equation}
\small
    \mW\;\in\;\R^{p\times q\times N_v\times N_h}, 
\quad
p = \Bigl\lceil\frac{m}{N_v}\Bigr\rceil,\;
q = \Bigl\lceil\frac{n}{N_h}\Bigr\rceil,
\end{equation}
so that each weight block \(W_{i,j} \in \mathbb{R}^{N_v\times N_h}\) is loaded onto a single PTC.

\begin{figure}
    \centering
    \includegraphics[width=0.8\columnwidth]{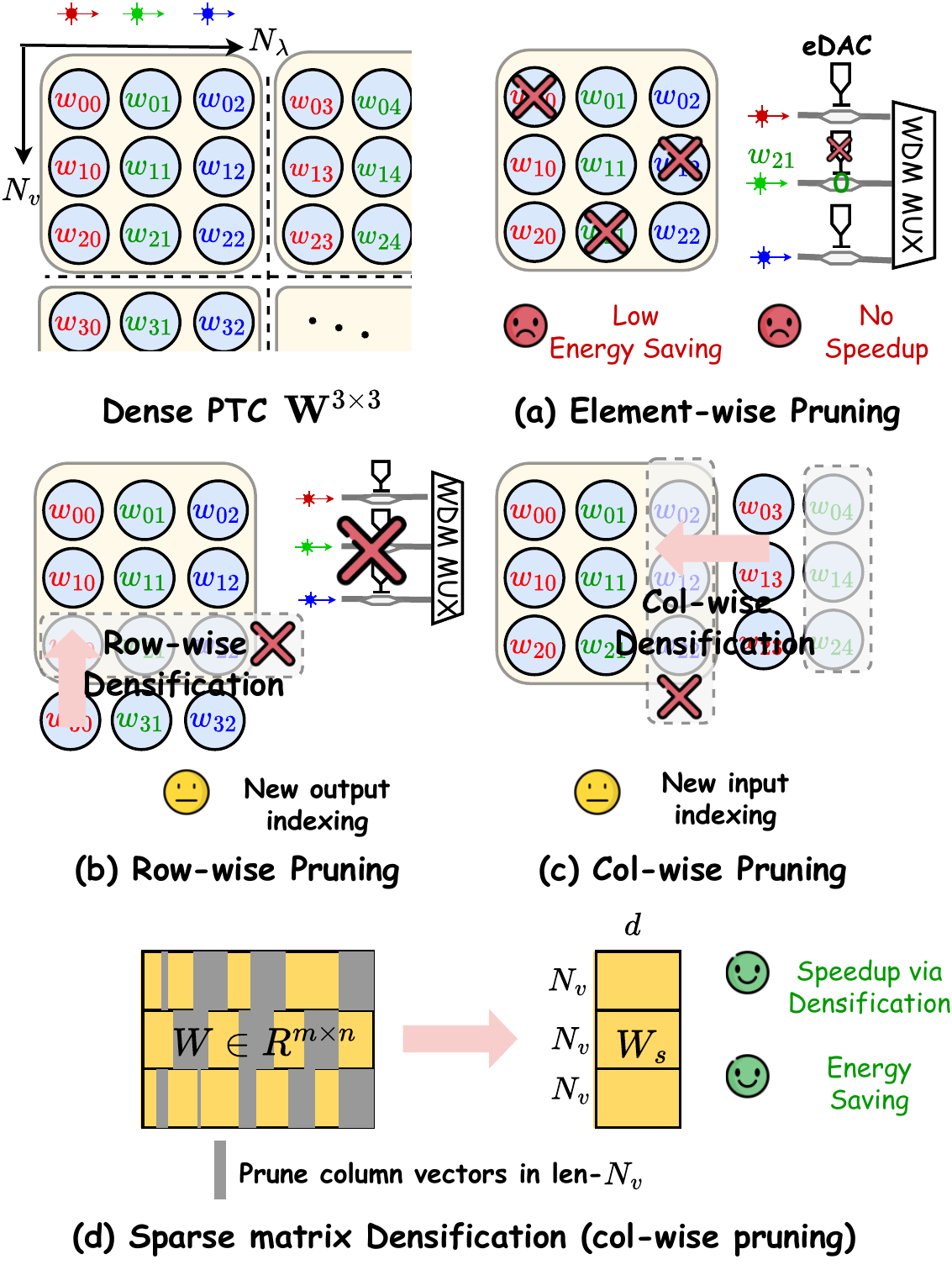}
    \vspace{-5pt}
    \caption{Starting from dense PTC, (a) unstructured element-wise sparsity provides no benefit due to dense computation. (b) Row pruning and (c) column pruning may not improve throughput and can lead to accuracy loss. (d) Our design balances energy savings, speedup, and performance.}
    \label{fig:SparseComparison}
    \vspace{-10pt}
\end{figure}

\underline{\textit{\textbf{Motivation:}}}
Photonic tensor cores execute tightly coupled vector–matrix operations.
Pruning the weight matrix in an unstructured manner requires encoding a logic 0, which still activates the modulation and driving circuitry  (see Fig.~\ref{fig:SparseComparison} (a)). 
Instead, hardware‑aware, structured sparsity  (e.g., pruning entire rows or columns) is more favorable (Fig.~\ref{fig:SparseComparison} (b) and (c)), enabling selective deactivation via further power gating or input gating.
However, naively applying structured sparsity may yield no speedup, as in a recent work~\cite{yin2024scatter} that reported no throughput benefits even aggressive 70\% compression.
Moreover, that work incurred over 7\% accuracy loss on CIFAR‑100 even after 200 epochs of sparse training.
We therefore seek sparsity patterns that not only match PTC constraints but also deliver \textit{energy savings} and \textit{speedup}.

\underline{\textit{\textbf{Our Solution:}}} \textit{\textbf{PTC-column Sparsity with Sparse Weights Densification}}:
Rather than pruning within each \(N_v\times N_h\) block \(W_{i,j}\)~\cite{yin2024scatter}, we apply column-wise structured sparsity in the size of PTC’s column granularity, along the horizontal dimension of the full weight matrix \(W\) (Fig.~\ref{fig:SparseComparison} (d)). 
After selecting and retaining the top \(d\) columns (by aggregate magnitude), we can shift those unpruned column-wise vectors left and “condense” each block into a smaller \(N_v\times d\) dense submatrix, reducing the overall weight footprint from \(m\times n\) to \(m\times d\).

We prefer column‑wise (over row‑wise) pruning for two reasons. 
First, \emph{flexible granularity}: column pruning naturally aligns with the PTC’s vertical dimension \(N_v\), and if finer pruning (\(<N_v\)) is required, we can apply row‑wise gating within each PTC to disable unused rows (see Sec.~\ref{subsec:reconfigurable_ptc}). 
Second, \emph{dataflow alignment}: column‑wise sparsity requires only input‑side indexing, which integrates seamlessly with the PTC’s output‑stationary dataflow~\cite{NP_HPCA2024_Zhu} to minimize ADC and buffer requirements. We adopt the lightweight input‑indexing design from~\cite{yang2019sparse}, incurring minimal area and power overhead.

\begin{figure}
    \centering
    \includegraphics[width=1.0\columnwidth]{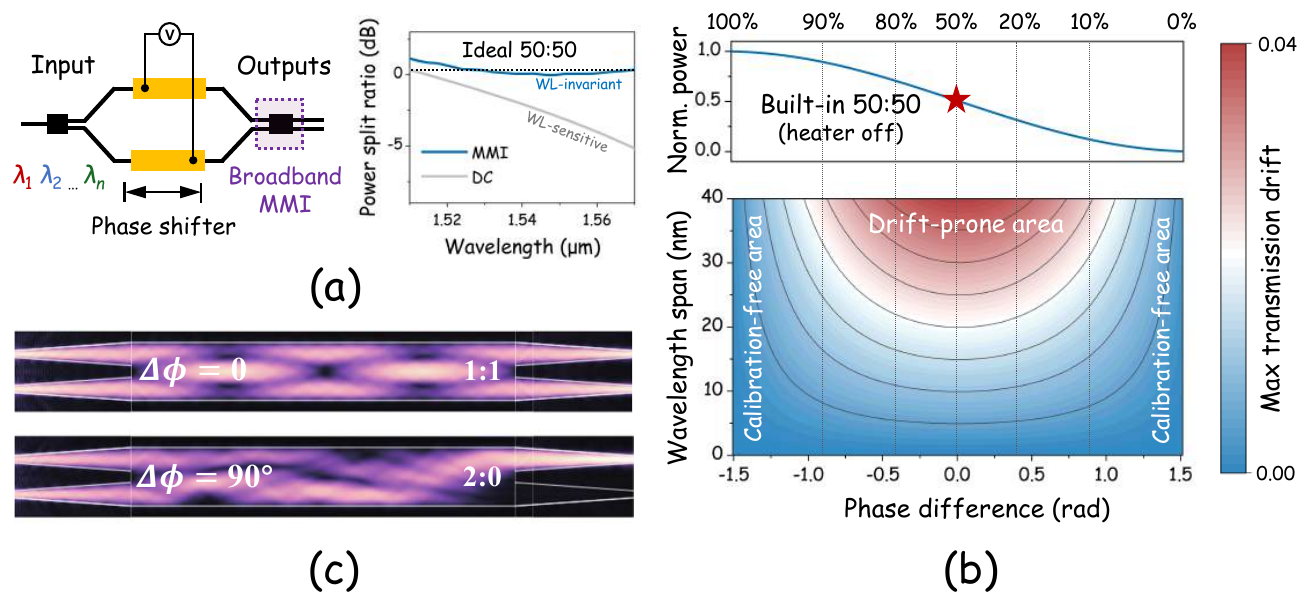}
    \vspace{-5pt}
    \caption{ (a) Broadband power redistribution unit using MMI for $\lambda$-tolerant splitting. (b)  Transmission drift across phase settings and total channel wavelength span. (c) Mode profile verification of equal and full power splitting scenarios.}
    \label{fig:hardware}
    \vspace{-5pt}
\end{figure}

\begin{figure*}
    \centering
    \includegraphics[width=\textwidth]{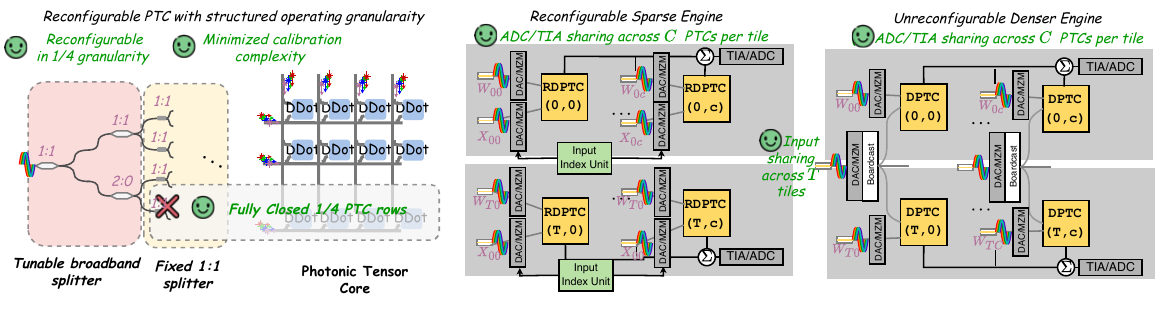}
    \vspace{-12pt}
    \caption{(a) Reconfigurable PTC with adaptive operating granularity; (b) Sparse engine with reconfigurable PTC (RPTC) and the corresponding input data fetcher to support our condensed sparse matrix; (c) Dense engine for uncompressed and low-rank factorized layers.}
    \label{fig:SPLIT-arch}
    \vspace{-5pt}
\end{figure*}

\subsection{Photonic Tensor Core with Reconfigurable Operating Granularity}
\label{subsec:reconfigurable_ptc}

\begin{figure}
    \centering
    \includegraphics[width=0.6\columnwidth]{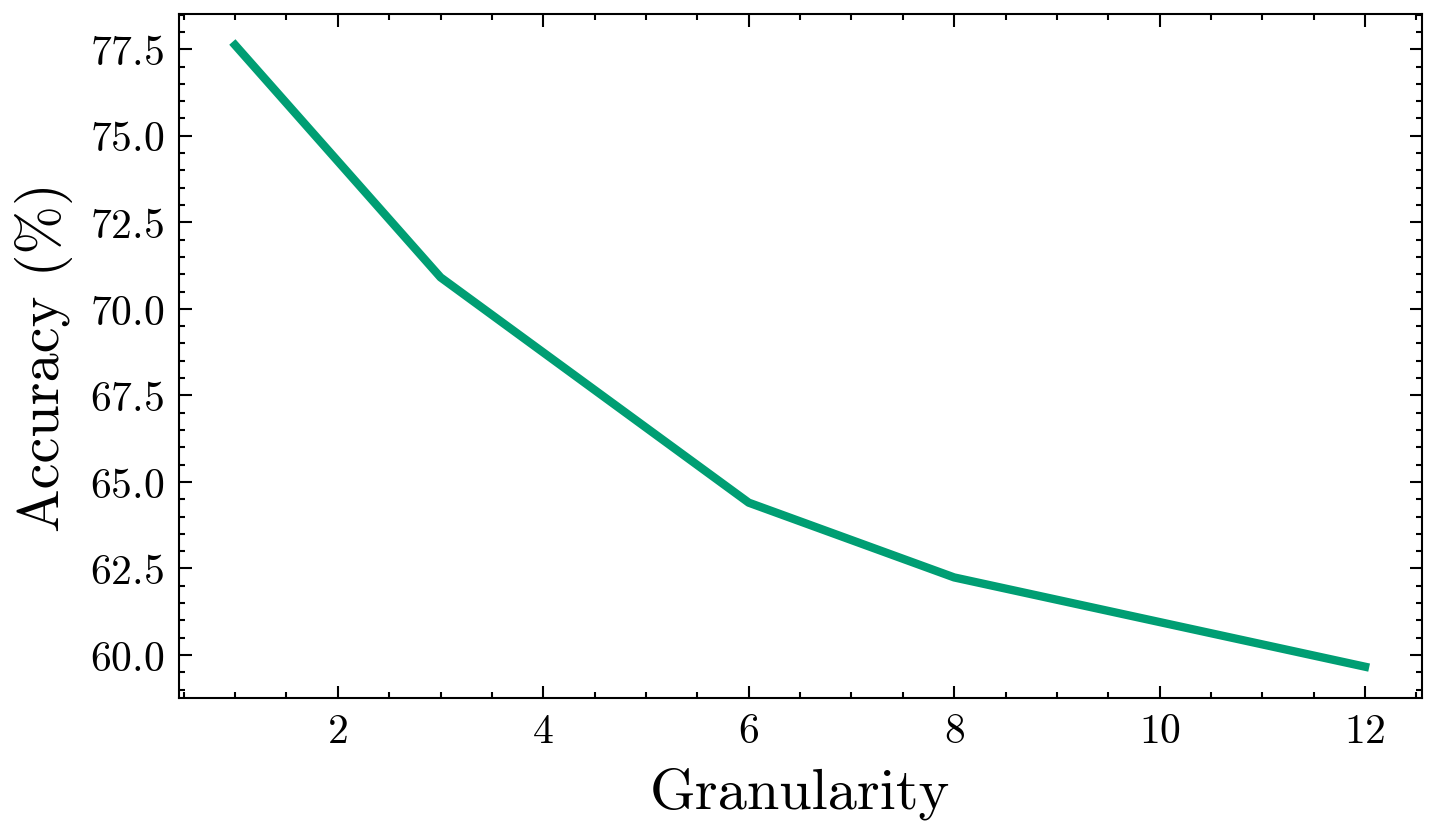}
    \vspace{-5pt}
    \caption{The Zer-shot compression accuracy of DeiT-Small (-40\%) under different compression granularities.}
    \label{fig:granularities}
    \vspace{-10pt}
\end{figure}

\underline{\textit{\textbf{Motivation:}}}  
Structured sparsity on fixed‑topology PTCs lacks post‑fabrication flexibility to adapt to varying pruning granularities. Different models and tasks demand different sparsity resolutions: for instance, zero‑shot compression accuracy on DeiT‑Small degrades steeply as the column‑pruning granularity grows from \(d=1\) to \(d=12\), as shown in Fig.~\ref{fig:granularities}. While coarser granularity can yield greater energy savings, it comes at the cost of accuracy.
Therefore, the accelerator should allow users to dynamically trade off between energy efficiency and model fidelity by reconfiguring the sparsity granularity after fabrication.

\underline{\textit{\textbf{Our Solution:}}} \textit{\textbf{Reconfigurable PTC with adaptive granularity}}: 
Inspired by SCATTER~\cite{yin2024scatter}, we design a dynamically reconfigurable PTC with an upstream tunable broadband optical redistributor. By steering the input optical paths, the PTC can switch among multiple operating granularities (Fig.~\ref{fig:SPLIT-arch} (a)), powering down inactive paths and eliminating wasted laser power for maximal energy savings.
However, this design is not directly transferable due to architectural and operational mismatches.
(1) Our PTC~\cite{NP_HPCA2024_Zhu} supports scalable wavelength-division multiplexed (WDM), which requires broadband-compatible components. In contrast, their design is restricted to single-wavelength operation. 
Without hardware-level redesign, wavelength-induced variations would incur significant inter-channel transmission deviations. 
(2)  Their approach is tightly coupled with a specific pruning scheme that demands arbitrary-ratio light reallocation (e.g., 5:1 or 7:2), which is infeasible in WDM systems. Due to inherent wavelength differences across WDM channels, each channel experiences distinct transmission characteristics, leading to substantial deviations from the intended power ratios. Compensating for these variations requires significant calibration overhead and complexity, both of which increase rapidly with system size.  These challenges necessitate a wavelength-insensitive light redistribution architecture that integrates seamlessly with our structured sparsity mapping and reduces implementation complexity.

\noindent\textit{\textbf{Key Design \ding{172} Broadband Calibration-Free Light Redistributor}}:
We adopt a reconfiguration scheme with 1/4-granularity, inspired by GPU 2:4 sparsity, by replacing only the first two splitter stages with broadband tunable splitters.
This setup requires only two discrete modes, 
full routing (2:0 or 0:2) and equal split (1:1), eliminating the need for arbitrary ratios. For example, as shown in Fig.~\ref{fig:SPLIT-arch} (a), after we switch one second‑stage splitter to 2:0 to suppress the input light, the corresponding 1/4 of PTC rows will be effectively disclosed.

To support WDM operation with no complex calibration, 
we introduce a broadband-optimized Mach–Zehnder interferometer (MZI) with multimode interferometers (MMIs) at the outputs instead of wavelength-sensitive directional couplers.
The two arms of the MZI are symmetrically designed, each incorporating a 200 microns thermal phase shifter (PS) to control power splitting.
By disabling the PSs, it defaults to a uniform 1:1 split, eliminating wavelength dependence arising from thermo-optic effects; at phase extremes it delivers full 2:0 or 0:2 switching with minimal wavelength sensitivity. This design ensures discrete, broadband-tolerant routing and true energy savings without complex calibration.

% In this configuration, disabling the PSs naturally yields equal splitting (1:1) , eliminating wavelength dependence arising from thermo-optic effects. When biased at the extrema of the phase response, the structure realizes full switching (2:0 or 0:2), where phase-induced transmission variations exhibit minimal wavelength sensitivity. This design thus satisfies the dual requirements of broadband tolerance and discrete routing support without per-channel calibration.  

\noindent\textit{\textbf{Simulation Validation}}: 
We performed full-wave FDTD device simulations and circuit-level S-parameter analysis (Fig. \ref{fig:hardware}b) to measure transmission drift across phase shifts. As expected, the maximum drift occurs at the linear point ($\Delta \phi = 0$) under wide WDM spans, but our symmetric design ensures equal splitting without tuning. Full-switching ($\Delta \phi = \pm \pi/2$) happens at the MZI’s extrema, where the phase response is flat and wavelength-insensitive. Mode-profile simulations (Fig. \ref{fig:hardware}c) confirm low-loss operation in both equal-split and full-switch modes. These results demonstrate our redistributor’s broadband, calibration-free performance.

\subsection{Overall \name Architecture}
\label{subsec:overall_arch}

Fig.~\ref{fig:SPLIT-arch} illustrates the high‑level organization of our \name accelerator, which comprises two complementary engines:

\textit{\textbf{Dense Engine}} inherits the design of~\cite{NP_HPCA2024_Zhu} to execute both standard dense matrix multiplications and low‑rank factorizations. It exploits photonic tensor cores (PTCs) operating at full resolution (\(N_v\times N_h\)).
% and benefits from global input broadcast to amortize the cost of optical modulation across multiple tiles.

\textit{\textbf{Reconfigurable Sparse Engine}} adopts the reconfigurable tensor core design discussed above with an adaptive operating granularity for varying sparsity granularity support.
Moreover, each tile handles each condensed chunk of the sparse matrix following the condensation process shown in Fig.~\ref {fig:SparseComparison} (d).
An input indexing unit is equipped to fetch the correct data, which we reuse the design presented in~\cite{yang2019sparse}, with low area and energy overhead.

\textit{\textbf{Optimization techniques:}}  
We further enable the following optimization techniques:
\ding{172} \emph{Photonic Input Broadcast:} For the dense engine, a global modulation unit drives multiple tiles in parallel, sharing the same input vector and thus reducing per‑tile DAC/modulation power.
 Sparse engines cannot broadcast (their inputs differ per tile), but because each sparse tile only processes a reduced input dimension (\(n \rightarrow d = n \times s\) with \(s=0.125\)), \textit{the additional encoding cost remains well-controlled};
\ding{173} \emph{Cross‑PTC ADC/TIA Sharing:} Both dense and sparse engines can employ an output‑stationary dataflow, as our PTC-column pruning with condensation preserves the output dimension.  Hence, we enable multiple PTC cores to share a single ADC/TIA bank by first accumulating in the analog domain.

\textit{\textbf{Scheduling Methodology:}}  
In this work, we adopt a simple static scheduler: low‑rank (dense) workloads are dispatched to the dense engine, while sparse workloads go to the reconfigurable sparse engine.
However, our sparse engines can also execute dense matrix multiplication, albeit at the cost of not broadcasting input data.
More sophisticated schemes—e.g., dynamic core partitioning between dense and sparse phases as in~\cite{you2023vitcod}—could yield additional \textit{latency improvements}, but are orthogonal to our core compression and hardware innovations.

\begin{figure*}
    \centering
    \includegraphics[width=0.85\textwidth]{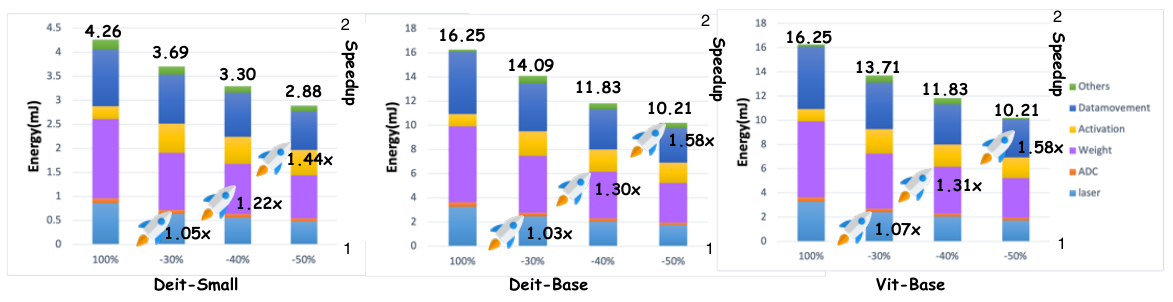}
    \vspace{-5pt}
    \caption{Energy and latency comparison between dense models and compressed models. We yield 2.5$\times$ energy-delay product reduction on Base-scale ViT under 50\% compression.}
    \label{fig:SparseLTNERGY}
    \vspace{-5pt}
\end{figure*}

\section{Experimental Results}
\label{sec:ExperimentalResults}
In this section, we evaluate the effectiveness of our compression flow \algname and the hardware efficiency of \name.
\subsection{Experiment Setup}
\label{sec:ExpSetup}

\noindent\textbf{Dataset and Models.}~  
We evaluate on the challenging ImageNet classification benchmark~\cite{deng2009imagenet} using three representative Vision Transformers: DeiT-Small, DeiT-Base, and ViT-Base. 
We focus on ViTs as their workload is \textit{compute-bound} with batched matrix multiplications, which is the sweet spot for the photonic AI engines.
LLM typically yields a sequential, autoregressive generation method, which is \textit{memory-bound}, not compute-bound~\cite{gao2024imi}. It is not yet suitable for photonics applications designed to accelerate computation.

\noindent\textbf{Decomposition and Training Settings.}~
We follow the same decomposition configuration in~\cite{zhang2025oats} by running the alternative SVD and pruning for 80 iterations.
We limit finetuning efforts to one hour (6 epochs for DeiT-small and 3 epochs for ViT-Base).
In our finetuning, the first epoch enables block-wise feature matching, while the remaining epochs focus on logit-level distillation.
\begin{figure*}
    \centering
    \includegraphics[width=0.85\textwidth]{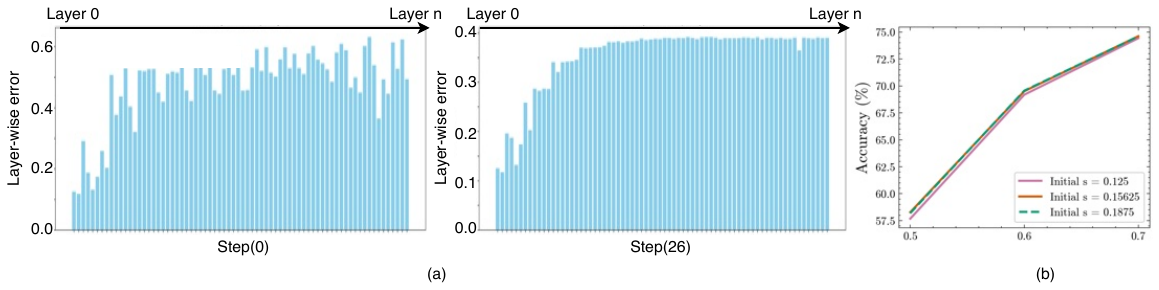}
    \vspace{-10pt}
    \caption{(a) The layer-wise error distribution change from Step(0) to the last Step(26); (b) The accuracy OF searched compressed model on varying initial sparsity ratios. 
    }
    \label{fig:Error}
    \vspace{-10pt}
\end{figure*}

\noindent\textbf{Hardware Settings.}~
\name builds upon the SOTA photonic Transformer accelerator~\cite{NP_HPCA2024_Zhu} but augments it with our new \emph{dynamically reconfigurable sparse engine}.
We use their \texttt{LT-Base} configuration with \(R=4\) tiles with \(C=2\) cores per tile. Each core is configured with a height of 12 and a width of 12, supporting 12 wavelengths. 
We add an extra 3 tiles of our new \emph{dynamically reconfigurable sparse engine}, wherein each core's height is set to 8. \emph{Note that our unique tensor core design enables operation at a \(1/4\) granularity to balance performance and accuracy.}  
To match peak compute capacity for fair comparison, we scale \texttt{LT-Base} with an extra 2 tiles of dense PTCs (with height of 12), denoted as \texttt{LT-Base-scaled}.
8-bit weights and 8-bit activation are used.
\emph{In this paper, we use a simple scheduling to allocate dense matrix multiplication onto dense engines, and sparse matrix multiplication on sparse engines. However, our sparse engines can also support dense matrix multiplication in full operating mode for additional speedup.}
We leave this for future study.
We use the simulator~\cite{NP_HPCA2024_Zhu} to model both computation and memory costs.

\subsection{Main Results}
\subsubsection{Evaluation of Our Compression Flow \algname}
Tab.~\ref{tab:imagenet_acc} reports top‑1 ImageNet accuracy under 30\%, 40\%, and 50\% parameter reductions. 
We compare against:  
(1) WANDA (sparse‑only)~\cite{sun2024a};  
(2) Truncated SVD (low‑rank only);  
(3) OATS~\cite{zhang2025oats}, in two variants—fixing the rank and solving for sparsity, or vice versa—under a uniform per‑layer budget.
We report both our zero‑shot compression results (\algnamea) and the accuracy after limited finetuning (\algnameb).

\noindent\textbf{Key Insights:} \ding{172} \textbf{Zero‑shot superiority:} Even without retraining, \algnamea consistently outperforms all baselines, closing the gap by up to 13\% on ViT‑Base compared to the best OATS variant, proving the effectiveness of our level 1 and level 2 solution in Sec.~\ref{sec:Method_alg}.
\ding{173}\textbf{Minimal fine‑tuning:} With just 3–6 epochs of distillation, \algnameb prunes DeiT‑Base and ViT‑Base by 50 \% with $\approx 1\%$ accuracy loss. On the smaller DeiT‑Small model, 30\% pruning incurs only a 1.5 point drop, which is acceptable since the smaller model naturally has less redundancy.
\ding{174}\textbf{Synergy of sparsity and low rank:} Combining structured sparsity with low‑rank factors significantly outperforms either technique alone; in particular, WANDA’s sparse‑only approach degrades severely under zero‑shot compression when honoring PTC-aware sparsity pattern.

These results demonstrate that our PTC‑aware decomposition flow, \algname, preserves model fidelity \textit{far better than state‑of‑the‑art baselines}, while \textit{respecting real photonic hardware constraints}.

\begin{table}[]
    \centering
    \caption{\small Top‑1 accuracy (\%) on ImageNet for various compression methods applied to DeiT and ViT. Results are shown at 30\%, 40\%, and 50\% parameter reduction. Sparsity granularity is set to 6 for DeiT-Small and 8 for others. \algnamea reports zero‑shot accuracy without retraining, and \algnameb reports accuracy after limited fine‑tuning.}
    \label{tab:imagenet_acc}
\resizebox{0.99\linewidth}{!}{%
\begin{tabular}{lcccc}
\toprule
 \multirow{2}{*}{\textbf{Params}} & \multirow{2}{*}{\textbf{Method}} & \multicolumn{2}{c}{\textbf{DeiT}} & \multicolumn{1}{c}{\textbf{ViT}} \\ 
\cmidrule(lr){3-5}
% &  & \textbf{Small ($\mathbf{22.1}\,\mathrm{M}$)} & \textbf{Base ($\mathbf{86.6}\,\mathrm{M}$)} & \textbf{Base ($\mathbf{86.6}\,\mathrm{M}$)} \\
&  & \textbf{Small} (22.1M) & \textbf{Base} (86.6M) & \textbf{Base} (86.5M) \\
\midrule
$-0\%$ & Original   & $79.80$ & $81.80$ & $80.99$ \\ 
\midrule 
\multirow{7}{*}{$-30\%$} 
   & Wanda               & 43.37 & 71.52    & 61.40 \\ 
   & Low-rank               & 59.58 & 77.82    & 66.89 \\ 
   & OATS (fixed $r$)   & 72.59 & 79.20 & 74.58  \\ 
   & OATS (fixed $s$) & 70.47 & 79.31 & 73.42  \\ 
 & \cellcolor{Highlight}\algnamea           & \cellcolor{Highlight}74.78 & \cellcolor{Highlight}81.10 & \cellcolor{Highlight}77.71 \\
 & \cellcolor{Highlight2}\algnameb           & \cellcolor{Highlight2}\textbf{78.24} & \cellcolor{Highlight2}\textbf{81.76} & \cellcolor{Highlight2}\textbf{80.76}\\
\midrule
\multirow{7}{*}{$-40\%$} 
   & Wanda               & 21.13 & 57.55    & 32.15 \\ 
   & Low-rank               & 44.45 & 74.58    & 57.76 \\ 
   & OATS (fixed $r$)   & 64.40 & 76.67 & 67.53 \\ 
   & OATS (fixed $s$) & 62.50 & 77.02 & 66.11   \\ 
 & \cellcolor{Highlight}\algnamea           & \cellcolor{Highlight}69.30  & \cellcolor{Highlight}79.92 & \cellcolor{Highlight}73.86 \\
 & \cellcolor{Highlight2}\algnameb           & \cellcolor{Highlight2}\textbf{76.94} & \cellcolor{Highlight2}\textbf{81.26}& \cellcolor{Highlight2}\textbf{80.22} \\
\midrule 
\multirow{7}{*}{$-50\%$} 
   & Wanda               & 6.29 & 26.20    & 3.14 \\ 
   & Low-rank               & 22.43 & 68.19    & 40.48 \\ 
   & OATS (fixed $r$)   &44.70 & 70.44 & 52.50\\ 
   & OATS (fixed $s$) &46.09 & $72.03$ & 52.49 \\ 
   & \cellcolor{Highlight}\algnamea           & \cellcolor{Highlight}56.18 & \cellcolor{Highlight}76.80 & \cellcolor{Highlight}65.21 \\
 & \cellcolor{Highlight2}\algnameb           & \cellcolor{Highlight2}\textbf{74.75} & \cellcolor{Highlight2}\textbf{80.70} & \cellcolor{Highlight2}\textbf{79.52} \\
\bottomrule
\end{tabular}
}
\vspace{-10pt}
\end{table}

\subsubsection{Evaluation of Compatibility with PTC Precision and Noise}
We validate that our compressed models add no extra quantization difficulty under realistic photonic tensor core precision constraints—8‑bit weights and 8‑bit activations—by applying post‑training quantization (PTQ) to both our zero‑shot and finetuned compressed variants using PTQ4ViT~\cite{yuan2022ptq4vit}. We use per‑channel weight quantization and tensor‑wise activation quantization; the per‑channel weight quantization integrates seamlessly with our output‑stationary dataflow via a single rescaling step at write‑back.

As shown in Table~\ref{tab:ModelQuant}, 8‑bit quantization of our compressed models results in minimal accuracy loss thanks to the Hessian-driven optimization~\cite{yuan2022ptq4vit}.
Furthermore, we inject 3\% random noise into both weight values and activation encodings, following \cite{NP_HPCA2024_Zhu}, and observe that accuracy degradation remains below 1\%, demonstrating that our compression flow does not amplify sensitivity to realistic PTC noises.

\begin{table}[]
\centering
\caption{\small~Accuracy of compressed models under PTC precision and noise.}
\label{tab:ModelQuant}
\resizebox{0.9\linewidth}{!}{%
\begin{tabular}{@{}lcc@{}}
\toprule
& \begin{tabular}[c]{@{}c@{}}Finetuned compressed\\ DeiT-Small (-30\%)\end{tabular}   & \begin{tabular}[c]{@{}c@{}}Zero-shot compressed\\ DeiT-Base (-30\%)\end{tabular}\\ \midrule
 & \textbf{78.24} & \textbf{81.10} \\
\midrule
\quad + 8-bit Q & 77.99 & 80.84 \\
\quad \quad + Noise & 77.47 & 80.45\\
\bottomrule
\end{tabular}
}
\vspace{-5pt}
\end{table}

\subsubsection{Evaluation of Hardware Efficiency of \name}
We next quantify the system‐level benefits of deploying our \algname‐compressed models on the \name accelerator, measuring both energy and latency on the photonic engine, as shown in Fig.~\ref{fig:SparseLTNERGY}.
We report the energy and latency speedup for DeiT‑Small, DeiT‑Base, and ViT‑Base under 30\%, 40\%, and 50\% parameter reductions compared to dense models running on \texttt{LT-Base-Scaled}.
We also report the breakdown of energy costs, specifically the data movement, weight E-O and D-A costs and laser energy costs.

\noindent\textbf{Key Insights:}
\ding{172}: \textbf{Significant energy and latency gains:}  At 50\% compression on ViT‑Base, \name achieves approximately 40\% energy savings and a 1.5× inference speedup compared to the uncompressed dense baseline.
\ding{173} \textbf{Alleviating photonic bottlenecks:}  
    Compression substantially reduces the data‑movement and electro‑optic conversion costs—previously the dominant energy sinks—validating the reduction of the key photonic overheads identified in Fig.~\ref{fig:Teaser}. Note that the activation encoding overhead increases since the sparse engine cannot use the input broadcast trick, but does not dilute the energy-saving benefits.
\ding{174 }\textbf{Scheduling considerations:}  
    Our simple static scheduler (dense workloads to the dense engine, sparse workloads to the sparse engine) limits achievable speedup on less compressed models.  Dynamic or mixed scheduling strategies could further improve latency in future work.

\begin{table}[]
\centering
\caption{\small~Ablation on design choices of \algname on final model accuracy under 50\% 
parameter reduction.}
\label{tab:ModelAblation}
\resizebox{0.98\linewidth}{!}{%
\begin{tabular}{@{}lcc@{}}
\toprule
Method & DeiT-Small (50\%)  & DeiT-Base (50\%)\\ \midrule
Best \baselinea & 46.09\% & 72.03\% \\ \midrule
\algname (Searched rank) & 53.42\% & 75.82\% \\
\rowcolor{Highlight}
\quad + local low rank adaptation & 57.90\% & 76.80\% \\ \midrule
\quad + Fine-tuning w/o KD (All epoch)  & 73.94\% & 80.11\% \\
\quad + Block-wise KD (All epoch) & 68.50\% & 79.50\% \\
\quad + Model-wise KD (All epoch) & 74.32\% & 80.60\% \\
\rowcolor{Highlight2}
\quad + \begin{tabular}[c]{@{}c@{}}Block-wise (1 epoch) $\rightarrow$ \\ Model-wise KD (Remaining epoch)\end{tabular} & \begin{tabular}[c]{@{}c@{}}\textbf{74.75\%} \\ (\textcolor{red}{$\uparrow 16.85\%$})\end{tabular} & \begin{tabular}[c]{@{}c@{}} \textbf{80.70\%} \\ (\textcolor{red}{$\uparrow 4.1\%$})\end{tabular} \\ 
\bottomrule
\end{tabular}
}
\vspace{-5pt}
\end{table}

\subsection{Ablation study}

\noindent\textbf{Q1: Effectiveness of Design Choices in \algname:}  
To isolate the impact of each component in our pipeline, we perform a step‑by‑step ablation (Table~\ref{tab:ModelAblation}) on both DeiT‑Small and DeiT‑Base under 50\% parameter reduction.
\textbf{Key insights:} \ding{202}{ Effectiveness of rank search and local adaptation:}  
Starting from the OATS baseline (uniform budget) yields only 46.1\% on DeiT‑Small and 72.0\% on DeiT‑Base.  
Replacing the uniform assignment with our fast batch‑wise rank search raises accuracy to 53.4\% (+7.3\%) and 75.8\% (+3.8 \%), respectively.  Adding the fast local low‑rank adaptation further improves zero‑shot performance to 57.9\% (+4.5\%) and 76.8\% (+1.0\%), confirming that per‑layer fidelity gains translate directly into higher task accuracy.
\ding{203}{Effectiveness of our two‑stage distillation finetuning recipes:}  
Our full two‑stage KD schedule—one epoch of block‑wise alignment followed by five epochs of model‑level distillation—achieves the best results: 74.75\% on DeiT‑Small and 80.9\% on DeiT‑Base.  This confirms that aligning intermediate representations before final logit matching maximizes accuracy recovery under a tight training budget.
In summary, each design choice—\textit{heterogeneous rank allocation, local low‑rank adaptation, and decoupled two‑stage KD}—contributes significant, complementary gains. 

\noindent\textbf{Q2: Effectiveness of the Fast Rank Assignment Algorithm}  
Fig.~\ref{fig:Error} (a) plots each layer’s reconstruction error before and after our batch‑wise greedy rank allocation. Initially, errors are highly skewed—some layers suffer very large approximation errors while others are barely affected. After applying our allocator, the per‑layer errors become much more balanced. 
This smoothing of the error profile ensures that no single layer dominates the degradation budget, leading to lower overall reconstruction error and, consequently, better end‑task accuracy. These results confirm that our fast rank assignment effectively redistributes the compression budget to where it is most needed.

\noindent\textbf{Q3: Robustness of Rank Search Across Sparsity Settings} 
We assess the sensitivity of our batch‑wise greedy rank allocator to different initial sparsity levels by evaluating three per‑layer sparsity budgets on DeiT‑Small: 12.5\%, 15.625\%, and 18.75\%—equivalently retaining 48, 60, and 72 columns out of the 384‑dimensional hidden size, each a multiple of the PTC column granularity (12). As shown in Fig.~\ref{fig:Error} (b), our allocator automatically adapts the per‑layer rank allocations for each sparsity setting, yet delivers consistently high top‑1 accuracy. This confirms that the greedy algorithm reliably identifies near‑optimal rank configurations regardless of the chosen sparsity budget.

\section{Conclusion and Future Work}
\label{sec:Conclusion}

We have introduced \name, the first hardware–software co–design for modern-scale Transformer inference on photonic tensor cores. Our \algname compression pipeline combines activation–aware low–rank factorization with PTC–aligned structured sparsity to prune ViT-Base by 50 \% with only $\sim$1 \% top-1 accuracy loss after three fine-tuning epochs. Integrated with \name’s adaptive tensor cores and calibration-free light redistribution, this co-design delivers a 2.5× improvement in energy–delay product over existing photonic Transformer accelerators on a Base-scale ViT.
This work represents the first attempt to explore scaling photonic hardware to modern-scale AI models from a hardware and software co-design perspective, paving the way for broader deployment of optics-based acceleration in advanced machine-learning workloads.

\newpage
\bibliographystyle{IEEEtran}
\bibliography{./ref/Top_sim,./ref/Top,./ref/NN,./ref/NP, ./ref/ICCAD}

\begin{thebibliography}{10}
\providecommand{\url}[1]{#1}
\csname url@samestyle\endcsname
\providecommand{\newblock}{\relax}
\providecommand{\bibinfo}[2]{#2}
\providecommand{\BIBentrySTDinterwordspacing}{\spaceskip=0pt\relax}
\providecommand{\BIBentryALTinterwordstretchfactor}{4}
\providecommand{\BIBentryALTinterwordspacing}{\spaceskip=\fontdimen2\font plus
\BIBentryALTinterwordstretchfactor\fontdimen3\font minus \fontdimen4\font\relax}
\providecommand{\BIBforeignlanguage}[2]{{%
\expandafter\ifx\csname l@#1\endcsname\relax
\typeout{** WARNING: IEEEtran.bst: No hyphenation pattern has been}%
\typeout{** loaded for the language `#1'. Using the pattern for}%
\typeout{** the default language instead.}%
\else
\language=\csname l@#1\endcsname
\fi
#2}}
\providecommand{\BIBdecl}{\relax}
\BIBdecl

\bibitem{NP_NATURE2017_Shen}
Y.~Shen, N.~C. Harris, S.~Skirlo \emph{et~al.}, ``Deep learning with coherent nanophotonic circuits,'' \emph{Nature Photonics}, 2017.

\bibitem{NP_Science2024_Xu}
Z.~Xu, T.~Zhou, M.~Ma, C.~Deng, Q.~Dai, and L.~Fang, ``Large-scale photonic chiplet taichi empowers 160-tops/w artificial general intelligence,'' \emph{Science}, vol. 384, no. 6692, pp. 202--209, 2024.

\bibitem{NP_HPCA2024_Zhu}
H.~Zhu, J.~Gu, H.~Wang, Z.~Jiang, Z.~Zhang, R.~Tang, C.~Feng, S.~Han \emph{et~al.}, ``Lightening-transformer: A dynamically-operated photonic tensor core for energy-efficient transformer accelerator,'' in \emph{IEEE International Symposium on High Performance Computer Architecture (HPCA)}, 2024.

\bibitem{NP_ACS2022_Feng}
C.~Feng, J.~Gu, H.~Zhu, Z.~Ying, Z.~Zhao \emph{et~al.}, ``A compact butterfly-style silicon photonic--electronic neural chip for hardware-efficient deep learning,'' \emph{ACS Photonics}, vol.~9, no.~12, pp. 3906--3916, 2022.

\bibitem{NP_NatureComm2022_Zhu}
H.~Zhu, J.~Zou, H.~Zhang \emph{et~al.}, ``Space-efficient optical computing with an integrated chip diffractive neural network,'' \emph{Nature Commun.}, 2022.

\bibitem{NP_SciRep2017_Tait}
A.~N. Tait, T.~F. de~Lima, E.~Zhou \emph{et~al.}, ``Neuromorphic photonic networks using silicon photonic weight banks,'' \emph{Sci. Rep.}, 2017.

\bibitem{NP_Nature2021_Xu}
X.~Xu, M.~Tan, B.~Corcoran, J.~Wu, A.~Boes, T.~G. Nguyen, S.~T. Chu, B.~E. Little, D.~G. Hicks, R.~Morandotti, A.~Mitchell, and D.~J. Moss, ``{11 TOPS photonic convolutional accelerator for optical neural networks},'' \emph{Nature}, 2021.

\bibitem{NP_Nature2021_Feldmann}
J.~Feldmann, N.~Youngblood, M.~Karpov, H.~Gehring, X.~Li, M.~Stappers, M.~L. Gallo, X.~Fu, A.~Lukashchuk, A.~Raja, J.~Liu, D.~Wright, A.~Sebastian, T.~Kippenberg, W.~Pernice, and H.~Bhaskaran, ``Parallel convolutional processing using an integrated photonic tensor core,'' \emph{Nature}, 2021.

\bibitem{touvron2022deit}
H.~Touvron, M.~Cord, and H.~J{\'e}gou, ``Deit iii: Revenge of the vit,'' in \emph{European conference on computer vision}.\hskip 1em plus 0.5em minus 0.4em\relax Springer, 2022, pp. 516--533.

\bibitem{ning2024photonic}
S.~Ning, H.~Zhu, C.~Feng, J.~Gu, Z.~Jiang, Z.~Ying, J.~Midkiff, S.~Jain, M.~H. Hlaing, D.~Z. Pan \emph{et~al.}, ``Photonic-electronic integrated circuits for high-performance computing and ai accelerators,'' \emph{Journal of Lightwave Technology}, 2024.

\bibitem{hurst2024gpt}
A.~Hurst, A.~Lerer, A.~P. Goucher, A.~Perelman, A.~Ramesh, A.~Clark, A.~Ostrow, A.~Welihinda, A.~Hayes, A.~Radford \emph{et~al.}, ``Gpt-4o system card,'' \emph{arXiv preprint arXiv:2410.21276}, 2024.

\bibitem{comanici2025gemini}
G.~Comanici, E.~Bieber, M.~Schaekermann, I.~Pasupat, N.~Sachdeva, I.~Dhillon, M.~Blistein, O.~Ram, D.~Zhang, E.~Rosen \emph{et~al.}, ``Gemini 2.5: Pushing the frontier with advanced reasoning, multimodality, long context, and next generation agentic capabilities,'' \emph{arXiv preprint arXiv:2507.06261}, 2025.

\bibitem{agarwal2025cosmos}
N.~Agarwal, A.~Ali, M.~Bala, Y.~Balaji, E.~Barker, T.~Cai, P.~Chattopadhyay, Y.~Chen, Y.~Cui, Y.~Ding \emph{et~al.}, ``Cosmos world foundation model platform for physical ai,'' \emph{arXiv preprint arXiv:2501.03575}, 2025.

\bibitem{cong2025e3d}
W.~Cong, Y.~Liang, Y.~Zhang, Z.~Yang, Y.~Wang, B.~Ivanovic, M.~Pavone, C.~Chen, Z.~Wang, and Z.~Fan, ``E3d-bench: A benchmark for end-to-end 3d geometric foundation models,'' \emph{arXiv preprint arXiv:2506.01933}, 2025.

\bibitem{cong2025can}
W.~Cong, H.~Zhu, P.~Wang, B.~Liu, D.~Xu, K.~Wang, D.~Z. Pan, Y.~Wang, Z.~Fan, and Z.~Wang, ``Can test-time scaling improve world foundation model?'' \emph{arXiv preprint arXiv:2503.24320}, 2025.

\bibitem{zhang2025oats}
\BIBentryALTinterwordspacing
S.~Zhang and V.~Papyan, ``{OATS}: Outlier-aware pruning through sparse and low rank decomposition,'' in \emph{The Thirteenth International Conference on Learning Representations}, 2025. [Online]. Available: \url{https://openreview.net/forum?id=DLDuVbxORA}
\BIBentrySTDinterwordspacing

\bibitem{sun2024a}
\BIBentryALTinterwordspacing
M.~Sun, Z.~Liu, A.~Bair, and J.~Z. Kolter, ``A simple and effective pruning approach for large language models,'' in \emph{The Twelfth International Conference on Learning Representations}, 2024. [Online]. Available: \url{https://openreview.net/forum?id=PxoFut3dWW}
\BIBentrySTDinterwordspacing

\bibitem{yin2024scatter}
Z.~Yin, N.~Gangi, M.~Zhang, J.~Zhang, R.~Huang, and J.~Gu, ``Scatter: Algorithm-circuit co-sparse photonic accelerator with thermal-tolerant, power-efficient in-situ light redistribution,'' \emph{arXiv preprint arXiv:2407.05510}, 2024.

\bibitem{zhu2022elight_aspdac}
H.~Zhu, J.~Gu, C.~Feng, M.~Liu, Z.~Jiang, R.~T. Chen, and D.~Z. Pan, ``Elight: Enabling efficient photonic in-memory neurocomputing with life enhancement,'' in \emph{2022 27th Asia and South Pacific Design Automation Conference (ASP-DAC)}.\hskip 1em plus 0.5em minus 0.4em\relax IEEE, 2022, pp. 332--338.

\bibitem{NP_PIEEE2020_Cheng}
Q.~{Cheng}, J.~{Kwon}, M.~{Glick}, M.~{Bahadori}, L.~P. {Carloni}, and K.~{Bergman}, ``{Silicon Photonics Codesign for Deep Learning},'' \emph{Proceedings of the IEEE}, 2020.

\bibitem{NP_NaturePhotonics2021_Shastri}
B.~J. Shastri, A.~N. Tait \emph{et~al.}, ``{Photonics for Artificial Intelligence and Neuromorphic Computing},'' \emph{Nature Photonics}, 2021.

\bibitem{zhu2022elight}
H.~Zhu, J.~Gu, C.~Feng, M.~Liu, Z.~Jiang, R.~T. Chen, and D.~Z. Pan, ``Elight: toward efficient and aging-resilient photonic in-memory neurocomputing,'' \emph{IEEE Transactions on Computer-Aided Design of Integrated Circuits and Systems}, vol.~42, no.~3, pp. 820--833, 2022.

\bibitem{zhu2022fuse}
H.~Zhu, K.~Zhu, J.~Gu, H.~Jin, R.~T. Chen, J.~A. Incorvia, and D.~Z. Pan, ``Fuse and mix: Macam-enabled analog activation for energy-efficient neural acceleration,'' in \emph{Proceedings of the 41st IEEE/ACM International Conference on Computer-Aided Design}, 2022, pp. 1--9.

\bibitem{gu2022adept}
J.~Gu, H.~Zhu, C.~Feng, Z.~Jiang, M.~Liu, S.~Zhang, R.~T. Chen, and D.~Z. Pan, ``Adept: Automatic differentiable design of photonic tensor cores,'' in \emph{Proceedings of the 59th ACM/IEEE Design Automation Conference}, 2022, pp. 937--942.

\bibitem{gu2022neurolight}
J.~Gu, Z.~Gao, C.~Feng, H.~Zhu, R.~Chen, D.~Boning, and D.~Pan, ``Neurolight: A physics-agnostic neural operator enabling parametric photonic device simulation,'' \emph{Advances in Neural Information Processing Systems}, vol.~35, pp. 14\,623--14\,636, 2022.

\bibitem{zhu2024pace}
H.~Zhu, W.~Cong, G.~Chen, S.~Ning, R.~Chen, J.~Gu, and D.~Z. Pan, ``Pace: Pacing operator learning to accurate optical field simulation for complicated photonic devices,'' \emph{Advances in Neural Information Processing Systems}, vol.~37, pp. 67\,535--67\,555, 2024.

\bibitem{NP_TCAD2022_Gu}
J.~Gu, C.~Feng, H.~Zhu \emph{et~al.}, ``Squeezelight: A multi-operand ring-based optical neural network with cross-layer scalability,'' \emph{IEEE Transactions on Computer-Aided Design of Integrated Circuits and Systems (TCAD)}, 2022.

\bibitem{gu2024m3icro}
J.~Gu, H.~Zhu, C.~Feng, Z.~Jiang, R.~T. Chen, and D.~Z. Pan, ``M3icro: Machine learning-enabled compact photonic tensor core based on programmable multi-operand multimode interference,'' \emph{APL Machine Learning}, vol.~2, no.~1, 2024.

\bibitem{liu2024spinquant}
Z.~Liu, C.~Zhao, I.~Fedorov, B.~Soran, D.~Choudhary, R.~Krishnamoorthi, V.~Chandra, Y.~Tian, and T.~Blankevoort, ``Spinquant: Llm quantization with learned rotations,'' \emph{arXiv preprint arXiv:2405.16406}, 2024.

\bibitem{NP_DATE2020_Gu}
J.~Gu, Z.~Zhao, C.~Feng, H.~Zhu, R.~T. Chen, and D.~Z. Pan, ``\text{ROQ}: A noise-aware quantization scheme towards robust optical neural networks with low-bit controls,'' in \emph{IEEE/ACM Proceedings Design, Automation and Test in Europe (DATE)}, 2020.

\bibitem{zhao2024galore}
J.~Zhao, Z.~Zhang, B.~Chen, Z.~Wang, A.~Anandkumar, and Y.~Tian, ``Galore: Memory-efficient llm training by gradient low-rank projection,'' \emph{arXiv preprint arXiv:2403.03507}, 2024.

\bibitem{zhu2024apollo}
H.~Zhu, Z.~Zhang, W.~Cong, X.~Liu, S.~Park, V.~Chandra, B.~Long, D.~Z. Pan, Z.~Wang, and J.~Lee, ``Apollo: Sgd-like memory, adamw-level performance,'' \emph{arXiv preprint arXiv:2412.05270}, 2024.

\bibitem{yuan2023asvd}
Z.~Yuan, Y.~Shang, Y.~Song, Q.~Wu, Y.~Yan, and G.~Sun, ``Asvd: Activation-aware singular value decomposition for compressing large language models,'' \emph{arXiv preprint arXiv:2312.05821}, 2023.

\bibitem{candes2011robust}
E.~J. Cand{\`e}s, X.~Li, Y.~Ma, and J.~Wright, ``Robust principal component analysis?'' \emph{Journal of the ACM (JACM)}, vol.~58, no.~3, pp. 1--37, 2011.

\bibitem{chen2021scatterbrain}
B.~Chen, T.~Dao, E.~Winsor, Z.~Song, A.~Rudra, and C.~R{\'e}, ``Scatterbrain: Unifying sparse and low-rank attention,'' \emph{Advances in Neural Information Processing Systems}, vol.~34, pp. 17\,413--17\,426, 2021.

\bibitem{jaiswal2024galore}
A.~Jaiswal, L.~Yin, Z.~Zhang, S.~Liu, J.~Zhao, Y.~Tian, and Z.~Wang, ``From galore to welore: How low-rank weights non-uniformly emerge from low-rank gradients,'' \emph{arXiv preprint arXiv:2407.11239}, 2024.

\bibitem{yu2023svd}
Z.~Yu and C.-S. Bouganis, ``Svd-nas: Coupling low-rank approximation and neural architecture search,'' in \emph{Proceedings of the IEEE/CVF Winter Conference on Applications of Computer Vision}, 2023, pp. 1503--1512.

\bibitem{yang2019sparse}
T.-H. Yang, H.-Y. Cheng, C.-L. Yang, I.-C. Tseng, H.-W. Hu, H.-S. Chang, and H.-P. Li, ``Sparse reram engine: Joint exploration of activation and weight sparsity in compressed neural networks,'' in \emph{Proceedings of the 46th International Symposium on Computer Architecture}, 2019, pp. 236--249.

\bibitem{you2023vitcod}
H.~You, Z.~Sun, H.~Shi, Z.~Yu, Y.~Zhao, Y.~Zhang, C.~Li, B.~Li, and Y.~Lin, ``Vitcod: Vision transformer acceleration via dedicated algorithm and accelerator co-design,'' in \emph{2023 IEEE International Symposium on High-Performance Computer Architecture (HPCA)}.\hskip 1em plus 0.5em minus 0.4em\relax IEEE, 2023, pp. 273--286.

\bibitem{deng2009imagenet}
J.~Deng, W.~Dong, R.~Socher, L.-J. Li, K.~Li, and L.~Fei-Fei, ``Imagenet: A large-scale hierarchical image database,'' in \emph{2009 IEEE conference on computer vision and pattern recognition}.\hskip 1em plus 0.5em minus 0.4em\relax Ieee, 2009, pp. 248--255.

\bibitem{gao2024imi}
B.~Gao, Z.~Wang, Z.~He, T.~Luo, W.-F. Wong, and Z.~Zhou, ``Imi: In-memory multi-job inference acceleration for large language models,'' in \emph{Proceedings of the 53rd International Conference on Parallel Processing}, 2024, pp. 752--761.

\bibitem{yuan2022ptq4vit}
Z.~Yuan, C.~Xue, Y.~Chen, Q.~Wu, and G.~Sun, ``Ptq4vit: Post-training quantization for vision transformers with twin uniform quantization,'' in \emph{European conference on computer vision}.\hskip 1em plus 0.5em minus 0.4em\relax Springer, 2022, pp. 191--207.

\end{thebibliography}
\end{document}